# Custom Arrayed Waveguide Gratings with Improved Performance

*Arthur C. van Wijk, Christopher R. Doerr, and B. Imran Akca*[*]*,

A. van W. Author 1, B. I. A. Author 3,

LaserLaB, Department of Physics and Astronomy, VU University Amsterdam, De Boelelaan 1081, 1081 HV Amsterdam, The Netherlands

C.R.D. Author 2
Aloe Semiconductor, 1715 Highway 35, Suite 304 Middletown, NJ 07748, USA

**Abstract:** Arrayed waveguide gratings (AWGs) are key optical components of various new applications in telecommunication, astrology, medical imaging, and spectroscopy. It is a very powerful integrated light dispersion technology with significant flexibility for tailoring its performance to the individual system needs of each application. There are several examples of custom AWG designs in the literature aiming for improved system performance. In this review, we will provide an overview of the available methods for improving the bandwidth, spectral resolution, and transmission function shape of AWGs. The working principle as well as the advantages and disadvantages of each method will be discussed.

## 1. Introduction

The arrayed waveguide grating (AWG) is a planar versatile light dispersion component with high accuracy, robustness, and design flexibility[1]. It has become an attractive component not only for telecommunication (e.g. multiplexer or demultiplexer)[2,3] but also for medical imaging[4-6], spectroscopy[7], and astronomy[8,9]. Optical loss, bandwidth, spectral resolution, transmission function shape, crosstalk, and polarization sensitivity are the most important performance parameters of an AWG spectrometer and for different applications, one or several of these parameters has a significant importance. For example, for optical imaging (e.g. optical coherence tomography, OCT) and spectroscopy applications (e.g. Raman spectroscopy) an AWG with a very large bandwidth together with high resolution is needed, whereas for telecommunication applications an AWG with flat-top transmission response with low crosstalk is more beneficial. Custom AWG designs have been developed to provide a high-performance AWG device while addressing the needs of each application[10-61]. Compared to computational spectrometers[62]— a rapidly growing field, custom AWGs can provide higher resolution and larger operation bandwidth. Moreover, the accuracy of computational spectrometers is proportional to the amount of training data that has to be well-characterized and well-labeled, and doing so could be quite time-consuming and difficult. In this sense, AWGs are still the choice of technology in many current applications. In this review article, we will provide an overview of both traditional and unconventional approaches for improving AWG performance on the bandwidth, spectral resolution, and transmission function shape. The advantages and disadvantages of each method on optical loss, footprint, crosstalk, and complexity of fabrication will be provided. The first section will briefly discuss the working principle and design parameters of AWG to acquaint the reader with the topic and in the following section; several methods for AWG performance improvement will be given.

## 2. Working principle and design parameters of AWG

An AWG consists of input/output waveguides, two focusing slab waveguides, and a phased array of multiple waveguides with a constant path length difference $\Delta L$ between neighboring waveguides. The operation of an AWG is briefly explained, referring to **Figure 1a.** Light from an input waveguide becomes laterally divergent in a first free propagation region (FPR) and illuminates the input facets of an array of waveguides with a linearly increasing length that leads to their typical curved shape. For a central design wavelength $\lambda_c$ the phase difference at the output facets of adjacent array waveguides is an integer multiple of $2\pi$. Since these facets are arranged on a circle, a cylindrical wavefront is formed at the beginning of a second FPR, which generates a focal spot at the central output waveguide. The phase shift caused by the length differences between arrayed waveguides is linearly dependent on wavelength, the resulting wavelength-dependent phase gradient implies a tilt of the cylindrical wavefront at the beginning of the second FPR, which causes the focal spot to shift to a different output waveguide with wavelength.

The fundamental design parameters of an AWG spectrometer are central wavelength ($\lambda_c = \frac{n_c \Delta L}{m}$), wavelength resolution ($\delta\lambda = \frac{n_s d^2}{mR}$), and free spectral range ($FSR = \frac{\lambda_c}{m}$, true only if the group index is equal to the index), where $n_s$ is the effective refractive index of the slab region, $n_c$ is the effective refractive index of the array waveguides, $\Delta L$ is the length difference between two adjacent waveguides in the grating, $d$ is the separation between the array waveguides, $\Delta x$ is the focal point deviation in the imaging plane from its central position (also known as the separation between output waveguides), $R$ is the slab radius, $m$ is the grating order, $i$ and $j$ are the number of the input and output waveguides, respectively, $\theta_i$ is the angle of the far-field diffracted light cone of the input waveguide, and is $\theta_o$ the acceptance angle of one of the output waveguides, and $\lambda$ is the wavelength. The derivations of these formulas are given in Ref. [3]. Obtaining high spectral resolution and large bandwidth at the same time is very challenging because this requires a large path length difference and a large number of arms, which increases the device size dramatically and results in performance degradation (e.g. increased loss and higher crosstalk). Methods for overcoming this limitation will be presented in the next sections.

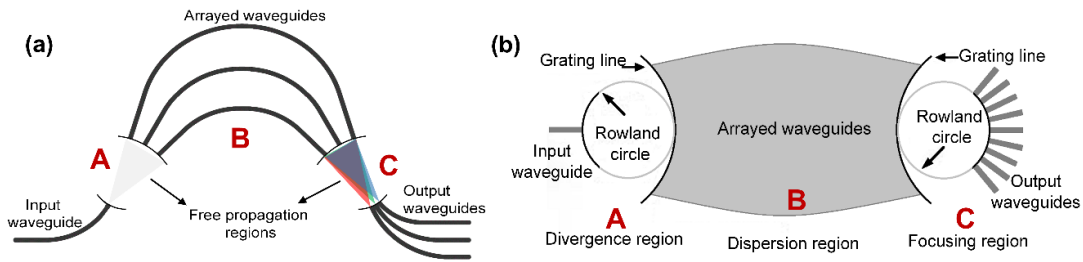

**Figure 1.** Schematic drawing of (a) a classical arrayed waveguide grating layout (AWG) and (b) different regions of an AWG. A, B, and C represent the divergence region (first free propagation region, FPR), dispersion region (arrayed waveguides), and focusing region (second FPR), respectively.

### 3. Performance improvement approaches

#### 3.1. Broad bandwidth operation

Combining high resolution and wide bandwidth (i.e. large FSR) in a single AWG is highly desirable for numerous applications, e.g. high-resolution OCT systems; however, it is rather difficult to implement. Several groups investigated different approaches to overcome the limitations on the resolution and FSR of an AWG[10-25]. The most common method is cascading several AWGs in a two-stage configuration, which enables the design of flexible AWG systems with high channel count and channel spacing. There are mainly two types of cascading configurations; "conventional" and "cyclic-FSR-based", which will be discussed in detail below.

### 3.1.1. Conventionally-cascaded AWGs

This method is based on the use of two stages of the conventional AWG structure with different resolutions and FSR values; the first stage, called the primary AWG, and the second stage, called secondary AWG, as illustrated in **Figure 2a**. This method is commonly used for high-density wavelength division multiplexing (WDM) systems; therefore, we call it the 'conventional' method.

The primary AWG, a 1 × N device, slices the whole spectrum of interest into wide N sub-passbands while the N secondary AWGs, 1 × M devices, reslice the wide passbands into narrow passbands. In this way, the overall configuration will have a wide FSR with narrow channel spacing (**Figure 2**). The center wavelength of each secondary AWG coincides with that of the corresponding primary filter passband. The main passband loss generated by the cascaded configuration consists of the loss at the primary and secondary filters. In addition to this loss, at the passband crossing points of the primary AWG, an extra loss will be introduced for the corresponding secondary AWG passbands due to the Gaussian-like transmission response of the AWG. Therefore, it is desirable to have a primary filter having a transfer function approximating a rectangular shape as much as possible. Different methods of realizing the rectangular transfer function (or flat-top response) will be discussed in the following subsections.

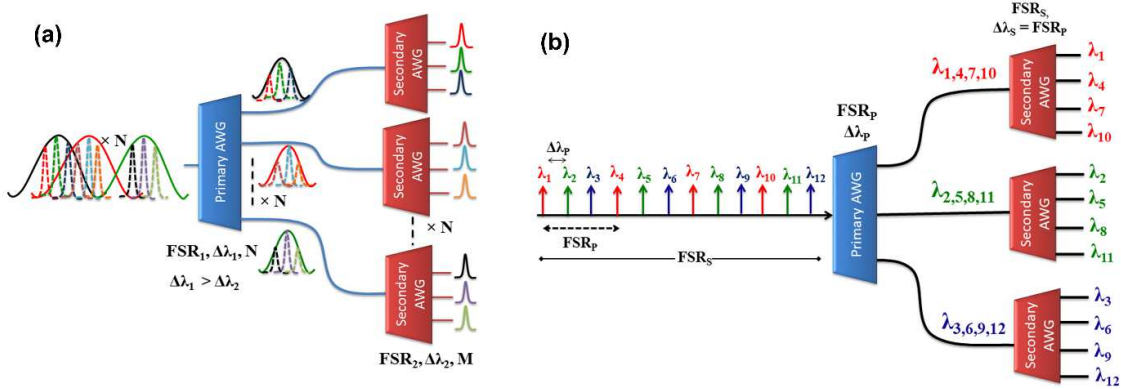

**Figure 2.** Schematic diagram of the (a) conventionally cascaded and (b) cyclic FSR nature-based AWG configurations. In (a) $N$ and $M$, respectively, indicate the number of primary and secondary filter output waveguides. In (b) $FSR_P$, and $FSR_S$ indicate the FSR of the primary and secondary AWGs, respectively, while $\Delta\lambda_P$ and $\Delta\lambda_S$ are the channel spacings of the primary and secondary AWGs, respectively. $\Delta\lambda_S$ equals $FSR_P$. The combined system has a free spectral range $FSR_S$ and a resolution $\Delta\lambda_P$.

There are several examples of two-stage tandem AWG structures in the literature[10-15]. Even though different bandwidths and spectral resolution values were demonstrated by different groups on different material platforms, the main working principle of each demonstration is the same and the improvement in each case mainly depends on the quality of the fabrication as well as the post-treatment of the devices for phase correction. In that sense, we will only discuss the cascaded AWG configuration with the largest FSR of 215 nm, which was achieved in a 25-GHz-spaced 1080-channel AWG using a tandem configuration[12] (see **Figure 3**). The non-adjacent crosstalk and loss values of that device ranged from -45 to -28 dB and 4.5 to 10 dB, respectively. However, these good crosstalk values were obtained after phase compensation based on a photosensitive phase-trimming technique[63,64], which involves irradiating the arrayed waveguide part with ultraviolet (UV) laser light through a metal mask that represents the phase errors (Fig. 3d). It is a universal method that can be applied to any material system. However, it complicates the fabrication of AWGs. Additionally, the major drawback of this design (as well as other conventionally cascaded AWGs) is the very large footprint of the device, which renders it impractical for most applications.

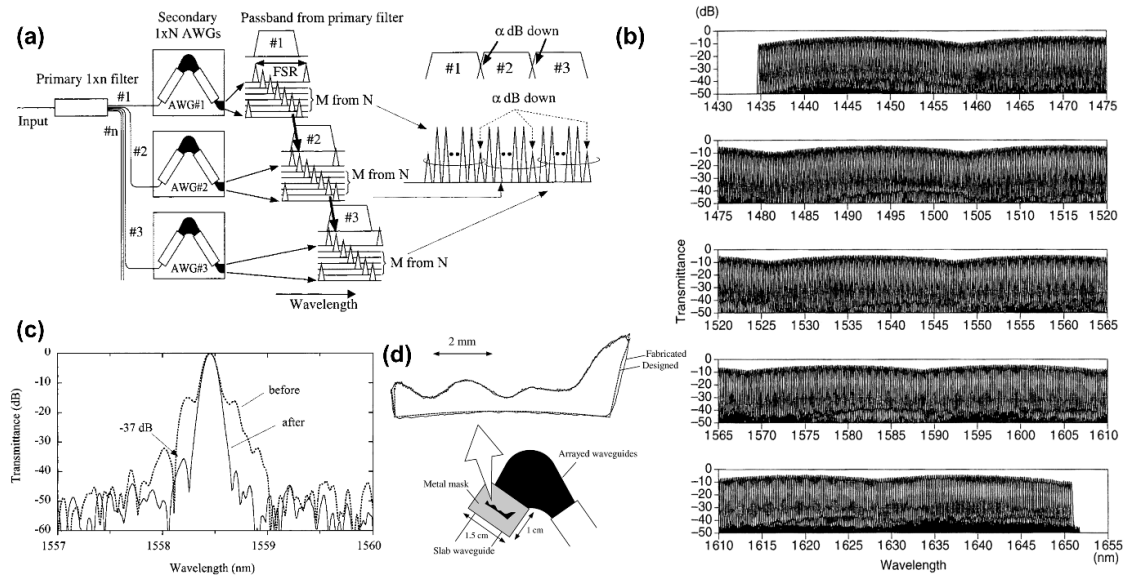

**Figure 3.** (a) Schematic diagram of tandem multi/demultiplexer. n and N, respectively, indicate the number of primary and secondary filter output ports. M is the number of selected output ports in the secondary AWG. (b) Transmission spectra of all 1080 channels of the tandem multi/demultiplexer. (c) Change in transmission spectra in the TE mode of one channel in a secondary 25-GHz-spaced AWG before and after photosensitive phase error compensation. Reproduced with permission.[12] Copyright 2002, IEEE. (d) Layout of an AWG with a metal mask for UV laser irradiation and comparison of window patterns in fabricated (solid line) and designed (dotted line) metal masks. Reproduced with permission.[64] Copyright 2005, IEEE.

A radically different cascaded AWG configuration that could provide the largest bandwidth was proposed by Nikbakht *et al.*[15] (see **Figure 4**). A wavelength range of 1200 nm (400 nm-1600 nm) with a 2 nm spectral resolution was numerically demonstrated. The overall device size is only $3 \times 3$ cm$^2$. The ultra-wide spectral range is made possible by using novel on-chip band-pass filters for the coarse wavelength division in contrast to previous work where another AWG was used as the primary filter. The fine resolution is provided by the AWG devices. The band-pass filter is formed by using bend waveguides and adiabatic full-couplers. There is no intrinsic loss associated with the band-pass filter, which makes the overall spectrometer loss relatively small. However, this design requires multiple inputs and there is no experimental demonstration yet.

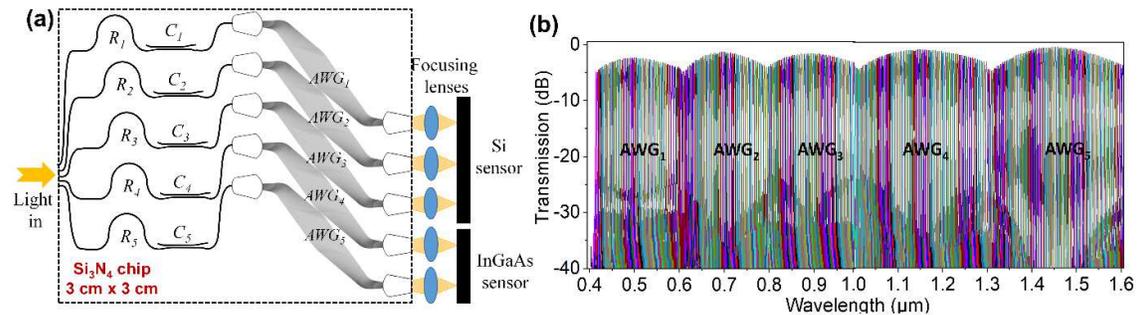

**Figure 4.** (a) Schematic of the ultrawide-bandwidth spectrometer using on-chip band-pass filters and AWGs. Outputs of AWGs are imaged onto Si and InGaAs sensors via focusing lenses. The rectangular

area indicated by black dashed lines shows the $Si_3N_4$ chip. (b) BPM simulation results of five AWG devices ($AWG_{1-5}$). Reproduced with permission.[14] Copyright 2020, OSA.

An interesting design was made by Jiang *et al.* in which they demonstrated a 4-AWG system fabricated on a dual-layer polymer platform[61] (**Figure 5**). Four different AWG designs can be mapped on a single chip by fabricating the upper and lower layers of the chip using the same photolithography mask that has been rotated 180 degrees. The measured transmission spectrum was 112 nm. Besides providing a large bandwidth, the footprint of this design is small; however, the alignment between the upper and lower layers could be a challenge.

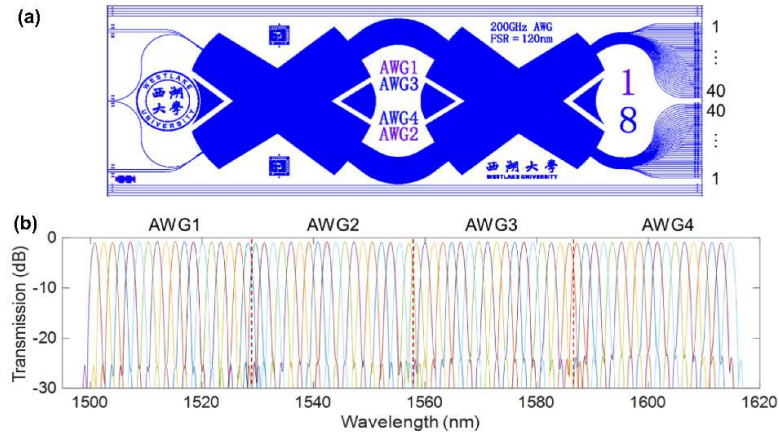

**Figure 5.** (a) Device layout of the dual-layer AWGs. (b) Simulated transmission spectra of four AWGs. Reproduced with permission.[61] Copyright 2021, OSA.

*3.1.2 Cascaded AWG systems based on cyclic FSR nature (Interleaved AWGs)*

An AWG has a cyclic transmission characteristic in that all wavelengths with a spacing equal to the FSR are transmitted simultaneously to a single output waveguide. Considering the size and loss problems of the conventional method described earlier, a new cascading technique with the use of this periodic routing property of the AWG can provide a much smaller system size and better overall performance (i.e. interleaved). In this configuration, instead of the primary AWG having wide flat passbands, it could have narrow closely spaced passbands (that equal the final desired channel spacing) that repeat *N* times in the desired wavelength range, using the frequency-cyclic nature of the AWG as illustrated in **Figure 2b**. The channel spacing of the secondary AWGs should be equal to the FSR of the primary AWG. In this configuration, the FSR of the secondary AWGs defines the FSR of the overall configuration whereas the channel spacing (resolution) of the primary AWG defines the overall system resolution. In this method, the requirements on resolution and passband flatness of the secondary AWGs are relatively low, as the components in their prefiltered input spectra can be easily separated with very low adjacent-channel crosstalk. Moreover, the complete device size will be much smaller compared to one that is designed with the conventional cascading method, since high-resolution AWGs are generally bigger than low-resolution AWGs so a system with one high-resolution AWG and multiple low-resolution AWGs can be smaller in overall size than one with a low-resolution AWG and multiple high-resolution AWGs. Using this method can be very beneficial for certain applications, e.g. OCT imaging, since the FSR of an AWG (thereby the wavelength resolution) is constant in frequency, which means that the spectrum is resolved in constant wavenumber $\Delta k$ intervals. Figure 2b illustrates the working principle. The channel spacing of the secondary AWGs, i.e. $\Delta\lambda_s$, should be equal to the FSR of the primary AWG, i.e. $FSR_P$. In this configuration, the FSR of the secondary AWGs, i.e. $FSR_S$, defines the FSR of the overall configuration whereas the channel spacing of the primary AWG, i.e. $\Delta\lambda_P$, defines the overall system resolution.

A theoretical analysis of multi-stage wavelength division multiplexing (WDM) networks was made by Maier *et al.* including interleaved AWGs[16]. Chan *et al.* demonstrated a hybrid multiplexer by combining interleaved AWGs with free space gratings using free space configuration[17]. The first experimental demonstration of cascaded AWGs based on cyclic FSR nature was done by Baier *et al.*[18] (see **Figure 6a**). They demonstrated a two-stage interleaved AWG spectrometer with a 1 nm resolution over 100 nm bandwidth in an indium phosphide (InP) platform. They also implemented photodiodes on the same chip (**Figure 6b**). The simulation and measurement results are given in **Figure 6c**. Although the optical loss is relatively high, the performance of the cascaded system was very good and in agreement with the simulation results.

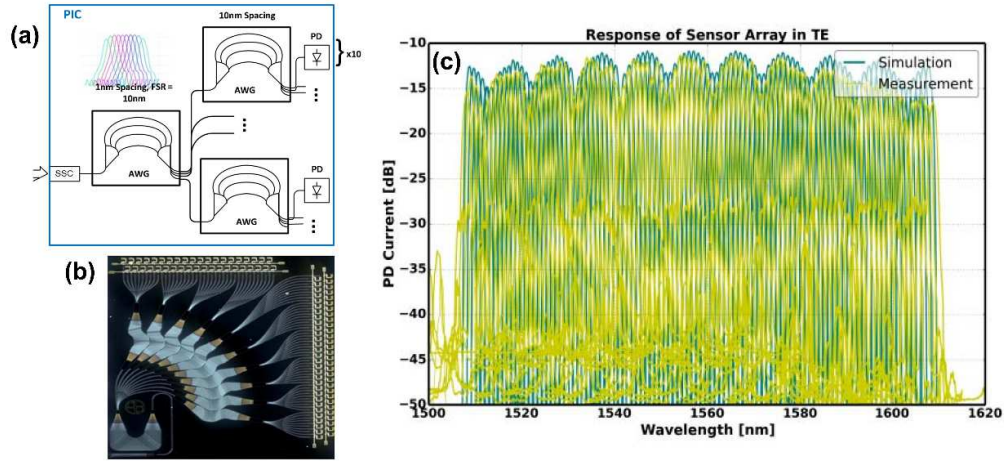

**Figure 6.** The design of the PIC (a) and a photo of the fabricated device (b). 11 AWGs of two different types are cascaded in two stages. 100 photodetectors (PD) are integrated behind the second stage. (c) Simulations and measurement data. Reproduced with permission.[18] Copyright 2014, ECOC.

Another experimental demonstration of a two-stage cyclic AWG system was provided by Akca *et al.* in a silicon nitride ($Si_3N_4$) platform with a 1 nm resolution over 75 nm bandwidth centered at the 1550 nm wavelength range[19] (**Figure 7**). A conventional AWG spectrometer with the same bandwidth and resolution values was realized in order to compare its performance with the interleaved AWG. Based on the measurement results, interleaved AWGs provide lower adjacent crosstalk values, and better channel uniformity, especially at the outer channels in addition to smaller device size. For a higher resolution and larger bandwidth, the size of the interleaved AWG demultiplexer becomes significantly smaller than a conventional AWG.

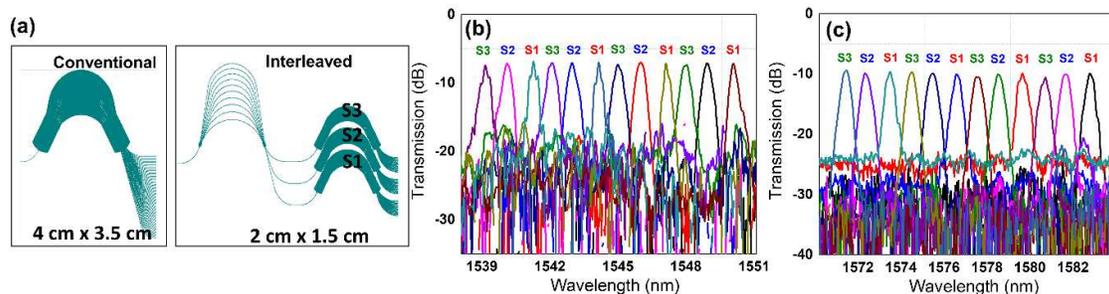

**Figure 7**. (a) Layout of the (left) conventional and (right) interleaved AWG spectrometers for the same resolution and bandwidth values. The size of the conventional design is ~ 2 times bigger than the interleaved design. Measurement results of the interleaved AWG for the (b) central and (c) outer waveguides. S1, S2, and S3 indicate the secondary AWGs as indicated in (a). Reproduced with permission.[19] Copyright 2019, IEEE.

Although the interleaved AWGs using the cyclic FSR nature promise much better overall performance compared to the conventional method, the fabrication of high-resolution AWGs can be very challenging because it requires a relatively large optical phase array. Moreover, this configuration is more sensitive to fabrication tolerances (i.e. thickness and refractive index nonuniformity of the core layer) than the conventional system for a high-resolution operation since the fabrication-related wavelength shift in the primary and secondary AWGs should match exactly in order to obtain the desired wavelength resolution. Micro-ring resonators (MRRs) have been proposed as an alternative to the high-resolution AWG to offer high spectral resolution with a small footprint[20-22]. Zhang *et al.* demonstrated a stationary spectrometer on a $Si_3N_4$ platform at the 860 nm center wavelength by using a tandem configuration of interleaved MRRs and multiple identical AWGs[23] (**Figure 8a**). The channel spacing of 0.75 nm and the optical bandwidth of 57.5 nm was achieved. However, precise control of multiple MRR resonances would necessitate either fine heater control of each MRR or additional fabrication techniques such as ion implantation and precise annealing to trim the specific position of each MRR's wavelength resonances[24]. In the following work of Zhang *et al.*[25] they demonstrated an alternative architecture of a scanning spectrometer that does not require precise trimming of the MRR. Their system is comprised of a scanning MRR as the primary stage and a single low-resolution AWG as the secondary stage, as illustrated in **Figure 8b.** By thermally tuning the resonance wavelength of the MRR, they demonstrated a cascaded spectrometer system with a 70-nm optical bandwidth and a 0.2-nm channel spacing. This is the widest bandwidth high-resolution integrated spectrometer in the 1300-nm wavelength region with a small footprint. A drawback of MRR's is their nonlinear group delay.

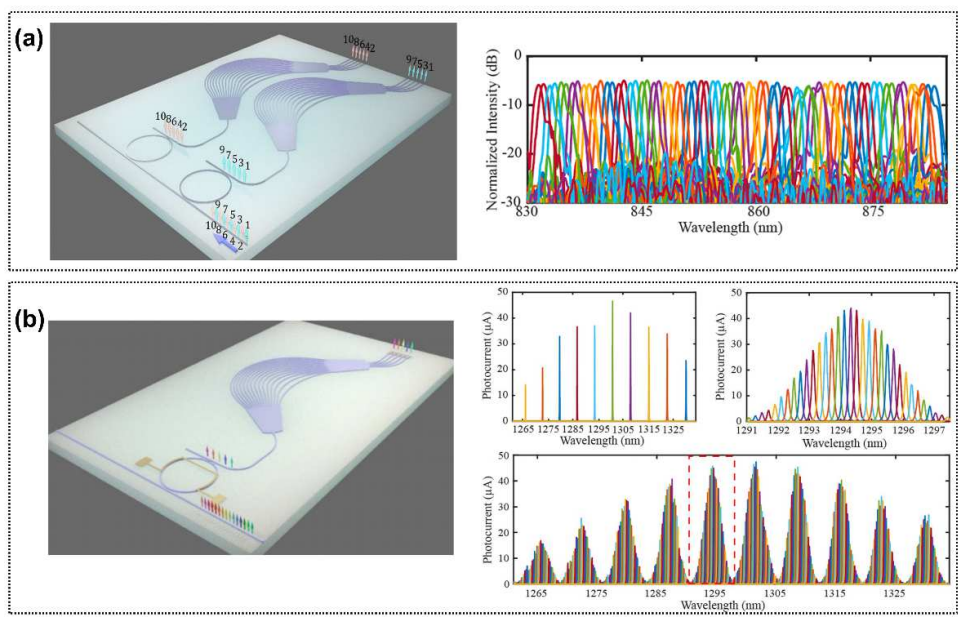

**Figure 8.** Schematic of the (a) stationary spectrometer with a pair of fixed wavelength MRRs and a pair of AWGs Reproduced with permission.[23] Copyright 2021, ACS, and (b) scanning spectrometer with tunable MRR and a single AWG. Measurement results of each type are given on the right side. Reproduced with permission[25] Copyright 2022, OSA.

### 3.3. Flat-top transmission response

The spectral response of a conventional AWG is Gaussian-like, but for many applications a flat spectral response is necessary. For example, the high loss and small passband due to the Gaussian transfer function of the primary AWG filters are still substantial problems of conventionally cascaded systems. An AWG with a flat transmission response can be achieved in several ways[26-43]. Takada *et al.* has demonstrated a partially integrated low-loss multiplexer/demultiplexer configuration by cascading bulky thin-film interference filters as primary filters and five AWGs as secondary filters in order to reduce the overall system loss[26]. Although the loss decreased considerably, the thin-film interference filters were still not integrated on-chip, which is a typical bottleneck for such filters. The commonly used approach is based on image mismatching between input and output waveguides. In this method, a double-peaked input field is convolved with the Gaussian profile of the output waveguides (see **Figure 9**). The main drawbacks of this approach are its inherent high loss and limited bandwidth.

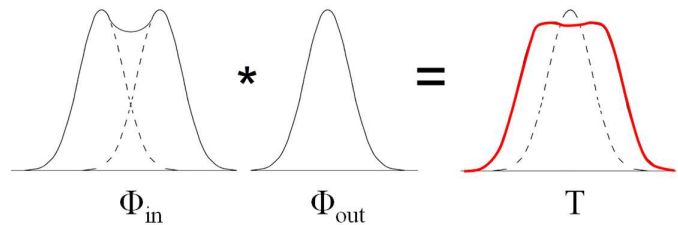

**Figure 9.** Image mismatching is a method in which two off-center images are created as the input function $\phi_{in}$. These are convoluted with the mode of the receiving waveguide ($\phi_{out}$ function) to form a broadened and flattened transmission T, at the cost of some loss.

The double-peaked input field can be created by
- Combining two waveguides, either output or input waveguides[30]
- Changing the shape of the input waveguides of the first FPR. Parabolic horns are the most widely used structure since they create interference between the first and second modes[31].
- Introducing an excess path length to the arrayed waveguides. It can be done either by inducing a phase difference at the outer arms[32] or creating a sync-like distribution of the spectrum at the output of the array[33].
- Using two AWGs with slightly different path length differences; i.e. $\Delta L$. For compactness, the AWGs are physically interleaved[34].
- Making the star couplers have two focal points[35] or a parabolic shape[36].
- Employing MMI couplers directly preceding the first star coupler[29, 37].

The flat-top passband is a highly desired feature, especially in cascaded AWG systems. Usually, a primary filter has a much smaller size than a secondary filter. A conventional cascaded AWG has a large footprint because it uses several big secondary AWGs. Moreover, due to the Gaussian transfer function, the insertion loss increases significantly at the passband crossings, which also increases the non-uniformity. One simple solution to this problem is to use fewer secondary AWGs with larger bandwidths, which also means that the primary filter has to have wider passbands preferably with a flat-top, i.e. rectangular, shape. In this sense, realizing primary AWGs with low-loss and box-like passbands over a broad spectral range is crucial. Such AWGs are also known as band demultiplexers, which are also very useful in both dense and coarse wavelength-division-multiplexed systems. Doerr *et al.* demonstrated a novel broad-band flat-top AWG design consisting of two waveguide grating routers, interconnected by a third one of non-standard design[44]. Despite the impressive device performance, the device size was $88 \times 5.4$ mm$^2$, which is rather large for further component integration, e.g. cascaded AWG systems (**Figure 10a**). Inoue *et al.* demonstrated a 20-nm-spacing 4-channel coarse wavelength-division-multiplexing filter, which comprises a total of 16 Mach-Zehnder interferometer (MZI) stages[45] (**Figure 10b**). Although the filter had the advantages of low loss, low crosstalk, and a wide passband, the device size was $15 \times 71$ mm$^2$. Wijk *et al.* demonstrated a compact cascaded AWG system based on an ultrabroad-spectral-range flat-top primary AWG[28] (**Figure 10c**). The broad bandwidth operation was provided by using an innovative multiple-input MMI coupler design. As a proof-of-principle, a 3-channel, 65-nm-spacing primary AWG was designed and a 0.5-dB bandwidth of 45 nm over 190 nm spectral range was demonstrated with an overall device size of 0.8 × 2.5 mm². A cascaded AWG system was realized in a Si$_3$N$_4$ platform using this flat-top 1×3 AWG as a primary filter and three 1×70 AWGs as secondary filters.

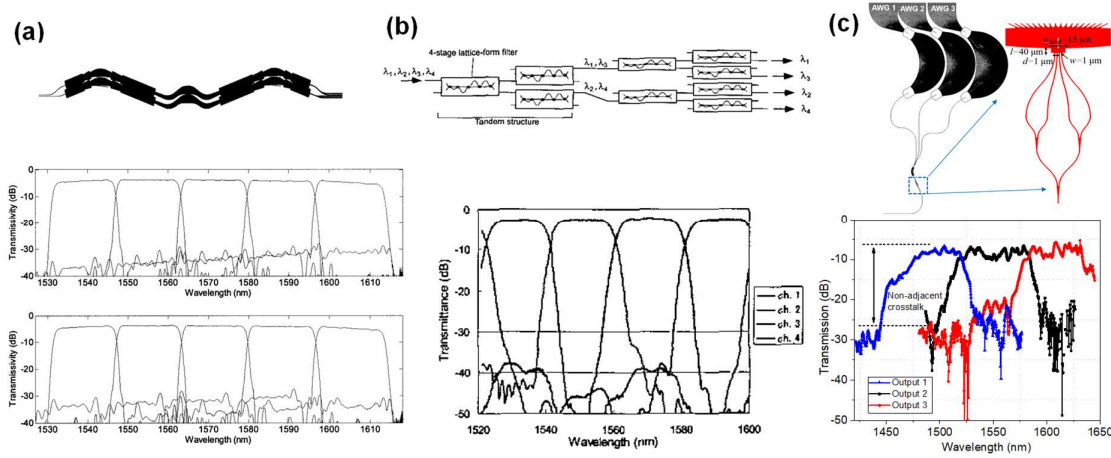

**Figure 10.** Schematic drawing (top) and measurement results of broad-band flat-top AWG design consisting of (a) two waveguide grating routers, interconnected by a third one of non-standard design[44], (b) 16 MZI stages[45] and (c) multiple-input MMI coupler at the first FPR[28]. Reproduced with permission.[44, 45, 28] Copyrights 2003, 2002, and 2020, IEEE, OFC, OSA, respectively.

Another method that has been widely used to achieve flat-top transmission response is the synchronized beam method, which was first described by Dragone[30] and later followed up by many groups[41,46,47]. This is a very effective method for increasing the passband width without introducing intrinsic loss. In most designs, an MZI-synchronized scheme is used in which the closely spaced dual-output waveguides of an asymmetrical MZI together form the input waveguides of the AWG, thus generating an input light spot that shifts with wavelength as illustrated in **Figure 11a**. The term "synchronized" refers to the FSR of the MZI being equal to the wavelength-channel spacing of the AWG, so that the input light spot for the AWG makes a full 'sweep' in position along the x-direction for each AWG wavelength channel. This synchronous movement of its input spot compensates the dispersion of the AWG for the wavelength range contained in one wavelength channel, thus imaging all these wavelengths at the same output waveguide. Since the FSR of the MZI is equal to the wavelength-channel spacing of the AWG, the resulting device has a flattened passband for all wavelength channels. The MZI configuration given in Refs. [30, 46, 47] was constructed by using a Y-junction coupler for the first coupler and a 2×2 directional coupler for the second coupler. The drawbacks of this configuration are excess loss and limited bandwidth due to the Y-junction coupler and the directional coupler, respectively.

In order to have a flat passband for all output channels, the channel spacing of the AWG should be equal to the FSR of the MZI, which is given as:

$$FSR_{MZI} = \frac{\lambda_c^2}{n_{eff} \Delta L_{MZI}}, \qquad (1)$$

where $\lambda_c$ is the center wavelength, $n_{eff}$ is the effective refractive index of the waveguide mode, and $\Delta L_{MZI}$ is the path length difference. Therefore, assuming that the MZI and arrayed waveguides have the same effective refractive index, the MZI should have a path length difference of

$$\Delta L_{MZI} = N \times \Delta L_{AWG}, \qquad (2)$$

where $N$ is the number of output channels of the AWG and $\Delta L_{AWG}$ is the constant length increment between adjacent arrayed waveguides.

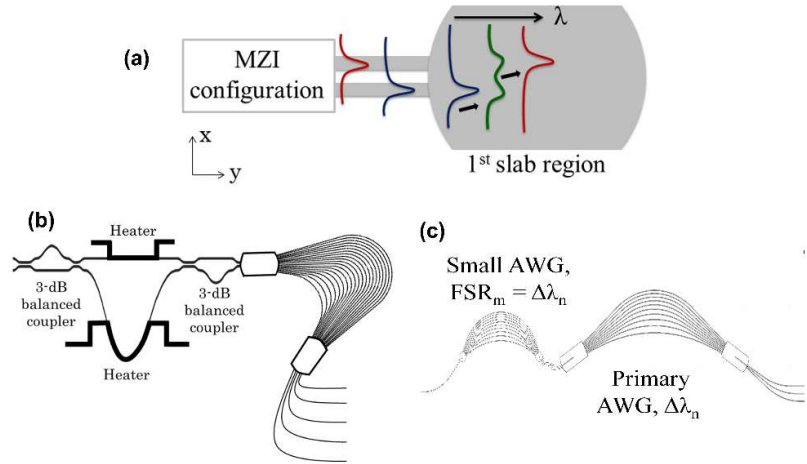

**Figure 11.** Synchronized beam method concepts. (a) Change in the optical field at the MZI and 1st slab region interface of the AWG, as the wavelength is changed. Reproduced with permission.[41] Copyright 2012, OSA. (b) Schematic of the broadband MZI-synchronized AWG using 3-dB balanced couplers. Reproduced with permission.[41] Copyright 2012, OSA. (c) Schematic of the low-loss flat-top primary AWG design, which uses a small AWG instead of MZI to achieve synchronized beam operation. Reproduced with permission.[28] Copyright 2020, OSA.

The broadband operation was also implemented in the synchronized beam scheme by using different methods[28,41]. In one approach, the directional coupler of the MZI is replaced with wavelength-insensitive 3-dB balanced couplers[41] (see **Figure 11b**). The phase deviation created by one of the balanced couplers is canceled by flipping the other coupler around. Using these components, a 5-channel, 18-nm-spacing MZI-synchronized AWG is designed and a 0.5-dB bandwidth greater than 12 nm over 90 nm spectral range is demonstrated with an overall device size of 11 × 3 mm$^2$. The relative phase difference between the two MZI arms due to the processing fluctuations is compensated by employing the thermo-optic effect to tune the synchronized MZI. The applicability of this approach is demonstrated by the design and realization of a cascaded AWG system, which consists of this flat-top 1×5 AWG as a primary filter and five 1×51 AWGs with a 0.4-nm channel spacing as secondary filters. A central excess-loss value of 4.5 dB (1 dB from the primary AWG and 3.5 dB from the secondary AWG) and a non-adjacent crosstalk value of 30 dB were obtained.

Another way of making a broadband flat-top AWG is using a small AWG at the first FSR[28], which is demonstrated in **Figure 11c**. A small AWG (i.e. 1.5 × 0.5 mm$^2$) with four closely spaced output waveguides is connected to the first FPR of the primary AWG in order to generate an input light spot that shifts with wavelength. The FSR of the small AWG, i.e. FSRm, is equal to the wavelength-channel spacing of the primary AWG, i.e. Δλn, to achieve synchronized beam operation. This method can provide broadband operation without introducing high optical losses as in the case of the aforementioned MMI-based design.

### 3.4 High-resolution operation

As explained in the earlier section, cascading AWGs can provide high resolution. However, there are also other unconventional methods that can improve the spectral resolution significantly[48-55]. One of these methods is removing the output waveguides of the AWG and re-imaging the output spectrum via a bulk lens on a pixel-based imaging array, e.g. CCD[49-52]. The overall spectral resolution of this configuration can be defined by the combination of arrayed waveguide(AW)-limited resolution and the CCD pixel size if the diffraction caused by the imaging lens is neglected. The convolution of these two factors yields the overall spectral resolution of the imaging system. The AW-limited resolution of an AWG can be explained by using **Figure 12**, which illustrates an optical imaging system comprised of two lenses with a focal length *R*, and a screen with an opening of width *Md*. It can be considered as a continuous equivalent of an AWG, with *M* the number of arrayed waveguides and *d* the spacing between them, except for removing all the sampling effects due to the discrete nature of the arrayed waveguides. According to Fourier analysis, the illumination at the output plane will be an image of the mode field profile of the input waveguide convolved with the sinc-type diffraction pattern of the screen opening.

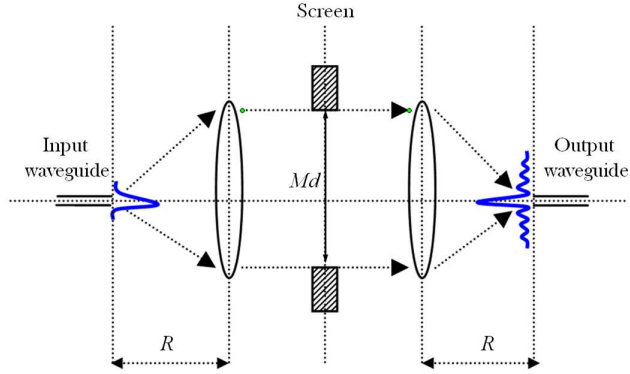

**Figure 12.** Continuous equivalent of an AWG spectrometer, with two lenses with a focal length *R*, and a screen with an opening of width *Md*.

The minimum angular FWHM of the output mode profile is determined by the Rayleigh criterion, which is given as:

$$\Delta\theta_{min} = \frac{\lambda}{Mdn_s}, \qquad (3)$$

where $n_s$ is the effective refractive index of the slab (lens) region. The angular dispersion of an AWG is given as:

$$\frac{\partial\theta}{\partial\lambda} = \frac{m}{dn_s}, \qquad (4)$$

where *m* is the grating order. From Eq. (4), $\Delta\theta$ can be extracted as:

$$\Delta\theta = \Delta\lambda \frac{m}{dn_s}, \qquad (5)$$

The relation between wavelength resolution and the number of AWs can be obtained by setting Eq. (3) equal to Eq. (5), which yields[56]:

$$\Delta\lambda_{AW} = \frac{\lambda}{Mm}, \qquad (6)$$

The absence of output waveguides is the most noticeable distinction between AWG-based astronomical spectrographs and telecom-type AWGs. AWGs without output waveguides have been widely used in astronomy to achieve higher resolution. **Figures 13 a,b and c** provide a conceptual sketch of the AWG-based spectrographs developed by different groups[9,51,50]. The output of the AWG is projected onto a camera via a bulk-optical system consisting of collimating and focusing optics as well as a blazed diffraction grating[9,51]. Because of the AWG's small FSR and cyclic behavior, which causes spectral orders to overlap, a (blazed) grating is used to separate the spectral orders, and expand the operation bandwidth of the AWG.

AWGs without waveguides have been used in biomedical applications as well in order to enhance the imaging range of an OCT system[49] (**Figure 13d**). The output spectrum was imaged onto a CCD using an ×10 objective lens, which increased the number of sampling points from 125 (number of output waveguides) to 450 (number of pixels addressed on the CCD). For the AWG with output channels, the wavelength resolution $\delta\lambda = 0.16$ nm leads to a theoretical maximum depth range in air of 1 mm. The overall spectral resolution i.e., the convolution of the diffraction pattern of the AW for a single frequency and the finite size of the CCD pixel, was calculated as 0.0465 nm. Experimentally, the maximum depth range was enhanced from 1 mm to 3.3 mm.

However, the curvature of the image plane (Rowland circle) causes defocus aberration for the outer output waveguides, which decreases the resolving power at off-axis wavelengths. Several methods of defocus correction are available, such as field flattening by integration of a lens into the second FPR[57], non-Rowland configurations utilizing a three-stigmatic-point method[58,59], and re-imaging of the spectrum using defocus-compensating bulk optics[60].

The packaging of AWGs without output waveguides is challenging due to the strict alignment requirements between the AWG chip, imaging lens, and the CCD camera. Usually, XYZ stages are used for fine alignment, which are costly components. The other problem could be the diffraction limit of the imaging lens whose effect could be reduced by increasing the numerical aperture of the lens. To the best of our knowledge, there is no commercial demonstration of a packaged AWG using a CCD camera; however, several groups demonstrated such a prototype as described in Refs. [9, 50,51].

Monolithic/on-chip integration of photodiodes[65,66] can ease the packaging issues related to fiber-chip outcoupling; however, customized electronics, control software, and suitable wire bonding technologies are required. Considering the large bandwidths and corresponding large number of output channels and integrated PDs, the implementation of customized electronics can be very challenging.

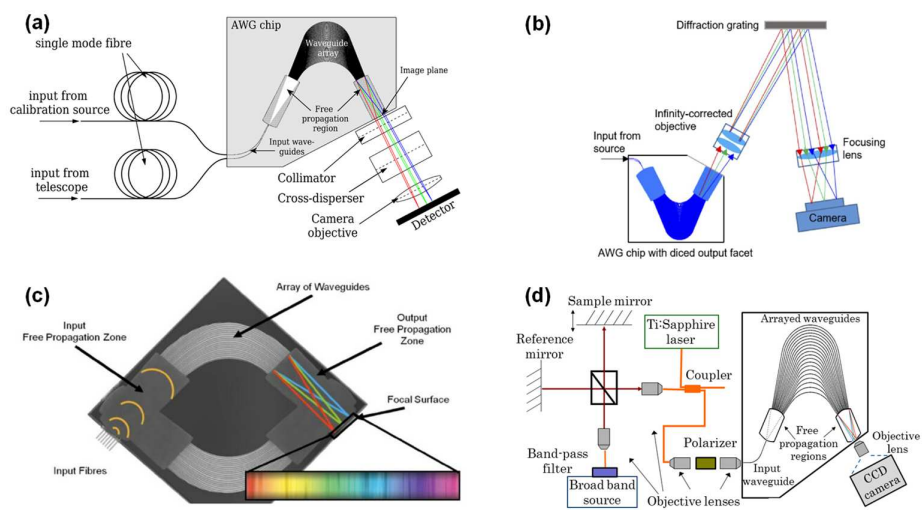

**Figure 13.** (a) Schematic depiction of the AWG-based spectrograph assembly with cross-dispersive optics[9]. (b) Conceptual layout of astronomical AWG-based spectrograph[51]. (c) Illustration of an AWG-chip configuration that has been modified for astronomical use by removing the output waveguides[50]. (d) Schematic of the experimental set-up of SD-OLCR with an AWG[49]. Reproduced with permission.[9,51,50,49.] Copyright 2017, 2020, 2012, 2012, MDPI, OSA, OSA, and IEEE, respectively.

Another method for improving the spectral resolution of an AWG is reusable delay lines[53]. Several directional couplers are used to distribute the input signal from the delay line into the FPR (**Figure 14a**). This approach reduces the footprint of the AWG significantly (71 times in this specific case) and therefore the impact of fabrication imperfections and nonuniformity can be minimized. Experimental results are given in **Figure 14b.** The optical loss and the crosstalk values are relatively high. One possible problem of this approach is the wavelength dependence of the directional couplers, which will limit the operating wavelength range of the AWG. However, using wavelength-insensitive couplers can solve this problem[67,68].

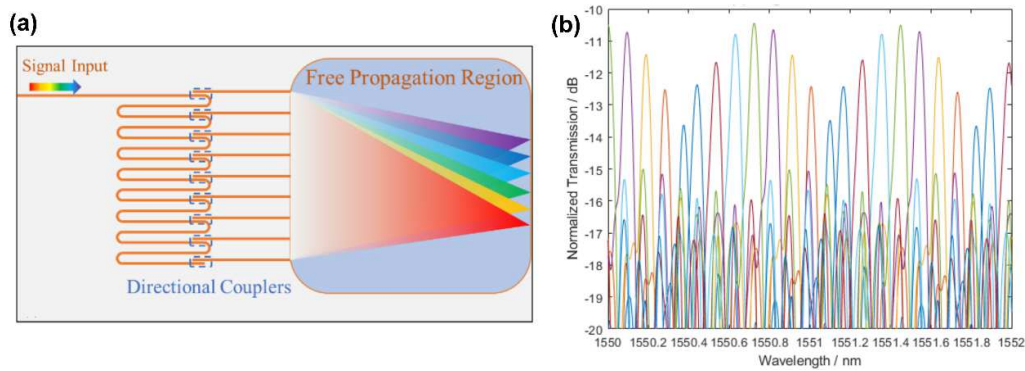

**Figure 14**. (a) A schematic of the AWG with reusable delay lines (RDL-AWG). (b) Experimental transmission response of the AWG with reusable delay lines. Reproduced with permission[53]. Copyright 2022, IEEE.

Sub-wavelength (SW) structures are another potential way for enhancing the spectral resolution of AWGs [54,55]. Despite the lack of a demonstration, this approach has the potential to give very high resolution with a very small footprint. An SW waveguide is a grating structure with a period shorter than the wavelength of light that suppresses diffraction and acts as a homogenous medium with an analogous refractive index. Compared to conventional optical waveguides, they offer an extra degree of freedom allowing tuning a few important waveguide properties, such as dispersion and refractive index. The arrayed waveguides of an AWG can be replaced with SW waveguides to achieve high resolution at a small footprint. The main challenges of this configuration are the high bend losses of the SW waveguides and the necessity of using more advanced fabrication methods to achieve subwavelength features. Moreover, it is very difficult to make a long-path-length difference using SW structures.

## 4. Conclusion

AWG is a powerful light dispersion technique and its performance can be tailored for each application by applying some traditional and unconventional methods as we explained in this review article. We also discussed the advantages and disadvantages of each method to provide a fair comparison between them. To have a large FSR, cascading several AWGs is the most commonly used method and there are different ways of cascading AWGs. Moreover, it is important to use a primary AWG with a flat-top and broadband range in order to avoid losses and reduce the footprint of the cascaded system. We provided an overview of different methods of realizing a flat-top as well as broadband transfer response. Cascading not only increases bandwidth but also helps to achieve high spectral resolution. In addition to that, we discussed other methods such as reimaging the second FPR on a large-pixel array, subwavelength gratings, and reusable delay line-based AWG designs. In this review, we have shown that by combining fundamental physical concepts with innovative approaches, it is possible to create very powerful, customizable, and high-performance AWG devices tailored to specific applications.

**Acknowledgments**

The authors received no financial support for the research, authorship, and/or publication of this article. The authors have no financial/commercial conflicts of interest.

Received: ((will be filled in by the editorial staff))

Revised: ((will be filled in by the editorial staff))

Published online: ((will be filled in by the editorial staff))

B. I. A. Author 2, A. van W. Author 2, C. R. D. Author 3, B. I. A. Corresponding Author*

**Custom Arrayed Waveguide Gratings with Improved Performance**

 In this review, we present that by combining fundamental physical concepts with innovative approaches, it is possible to create powerful, customizable, and high-performance AWG devices tailored to specific applications.